\documentclass[a4paper]{jpconf}
\usepackage{graphicx}
\usepackage[T1]{fontenc}
\usepackage[squaren]{SIunits}
\usepackage{amsmath,amssymb}

\newcommand{\YRS}{YbRh$_2$Si$_2$}
\newcommand{\LRS}{LuRh$_2$Si$_2$}
\newcommand{\SdH}{Shubnikov-de Haas}
\newcommand{\mK}{\milli\kelvin}
\newcommand{\mT}{\milli\tesla}
\newcommand{\kT}{\kilo\tesla}
\newcommand{\Freq}[1]{\unit{#1}{\kT}}

\begin{document}
\title{\SdH\ measurements on \LRS}

\author{S Friedemann\textsuperscript{\textsf{\bfseries 1}},
			S K Goh\textsuperscript{\textsf{\bfseries 1}},
			F M Grosche\textsuperscript{\textsf{\bfseries 1}},
			Z Fisk\textsuperscript{\textsf{\bfseries 2}},
			M Sutherland\textsuperscript{\textsf{\bfseries 1}}}

\address{\textsuperscript{\textsf{\bfseries 1}}{Cavendish Laboratory, University of Cambridge, JJ Thomson Avenue, CB3 0HE Cambridge, United Kingdom}
\emph{\textsuperscript{\textsf{\bfseries 2}}{Department of Physics and Astronomy, University of California, Irvine, CA 92697-4575, USA}}
}

\ead{sf425@cam.ac.uk}

%
%
\begin{abstract}
We present \SdH\ measurements on \LRS, the non-magnetic reference compound to the prototypical heavy-fermion system \YRS. We find an extensive set of orbits with clear angular dependences. Surprisingly, the agreement with non-correlated band structure calculations is limited. This may be related to an uncertainty in the calculations arising from a lack of knowledge about the exact Si atom position in the unit cell. 
The data on \LRS\ provide an extensive basis for the interpretation of measurements on \YRS\ indicative of discrepancies between the high-field Fermi surface of \YRS\ and the ``small'' Fermi surface configuration.
\end{abstract}

%
%
\section{Introduction}
Interest in the intermetallic compound \LRS\ is related to its relevance as a reference compound for the heavy-fermion system \YRS\ \cite{Rourke2008,Friedemann2010b}. In addition, \LRS\ turned out to feature  peculiar anomalies in the Hall effect \cite{Friedemann2010c}. For both purposes a good understanding of the electronic structure of \LRS\ is advantageous.

\YRS\ features a quantum critical point (QCP), i.e., a continuous phase transition at zero temperature. This QCP is accessed at small magnetic fields (\unit{60}{\mT} for fields perpendicular to the crystallographic $c$ direction). The QCP in \YRS\ appears to be a rare case  of an unconventional scenario \cite{Si2001,Coleman2001,Senthil2003}. The basic characteristic of such an unconventional QCP is that the entire Fermi surface becomes critical as part of the electrons become localized when tuning through the QCP. In heavy-fermion systems this partial localization is associated with the breakdown of the Kondo effect, with the composite quasiparticles formed through the Kondo effect disintegrating into their constituents, $f$- and conduction electrons. Evidence for the change of the Fermi surface is seen in  measurements of the Hall effect across the QCP \cite{Paschen2004,Friedemann2010b}. The Hall coefficient passes through a crossover at finite temperatures which sharpens to a discontinuity in the extrapolation to zero temperature with the values on the low and high field side of the QCP reflecting the ``small'' and ``large'', conduction electron and composite quasiparticle Fermi surface, respectively. 

The picture of the Fermi-surface reconstruction has been backed by de Haas-van Alphen (dHvA) studies on \YRS\ 
\cite{Rourke2008}. Although the observed angular dependence of the dHvA frequencies at fields above \unit{14}{\tesla} are best described by the ``small'' Fermi surface picture the increase of the dHvA frequencies below \unit{10}{\tesla} is interpreted in terms of the large Fermi surface. The transition at \unit{10}{\tesla} is speculated to be of Lifshitz type \cite{Rourke2009}. A large Fermi surface configuration as concluded from the dHvA measurements below \unit{10}{\tesla} is consistent with the Hall effect measurements indicative for a large Fermi surface configuration above \unit{0.06}{\tesla}. 
In later dHvA studies a higher frequency of $F = \unit{14}{\kT}$  was observed at  fields above \unit{13}{\tesla} \cite{Sutton2010}. This frequency was associated with the hole-like Fermi surface sheet of the ``small'' Fermi surface configuration consistent with the other frequencies. However, it remains an open question why no other orbit of this sheet could be observed. In addition, the agreement of the observed angular dependences with calculated orbits of either the ``small'' or ``large'' Fermi surface configuration are limited. Consequently, a better understanding of the ``small'' Fermi surface may help to interpret the dHvA measurements on \YRS.


Here, we present measurements of the \SdH\ (SdH) effect on \LRS, thereby providing an experimental reference for the ``small'' Fermi surface of \YRS. The heavy-fermion behaviour of \YRS\ is based on the magnetic moment provided by an incompletely filled 4$f$ shell. In \LRS\ the 4$f$ shell is completely filled and thus \LRS\ may serve as a non-magnetic reference to \YRS. In fact, the comparison can be extended to a quantitative level: \LRS\ features identical lattice parameters within experimental resolution which means that not only the topology but also the absolute dimensions of the Fermi surface of \LRS\ should resemble the ``small'' Fermi surface of \YRS. Consequently, \LRS\ is a very powerful reference system to \YRS.


%
%
\section{Experimental setup}
Single crystal samples were grown in indium flux as described earlier \cite{Maquilon2007}. They grow preferentially in form of thin ($\lesssim \unit{30}{\micro\meter}$) platelets which makes them ideal for SdH measurements. SdH oscillations were measured using a standard four probe ac resistivity technique. Contacts were provided by \unit{25}{\micro\meter} gold wires spot welded to the sample and stabilized with silver epoxy The sample size was approximately $0.02 \times 0.1 \times \unit{2}{\cubic\milli\meter}$. The current was applied within the basal plane at an angle of $\approx \unit{10}{\degree}$ from the (100) axis. Measurements were performed in a $^3$He/$^4$He-dilution refrigerator in fields up to \unit{16}{\tesla}. The oscillatory part was deduced by subtraction of a 4\textsuperscript{th} order polynomial fit from the raw data. Oscillation frequencies were determined after Fourier transformation. In order to deduce the angular dependence of the oscillation frequencies the magnetic field was rotated within the crystallographic basal plane.

%
%
\section{Experimental Results and Discussion}
\begin{figure}%
\centering
\includegraphics[width=\textwidth]{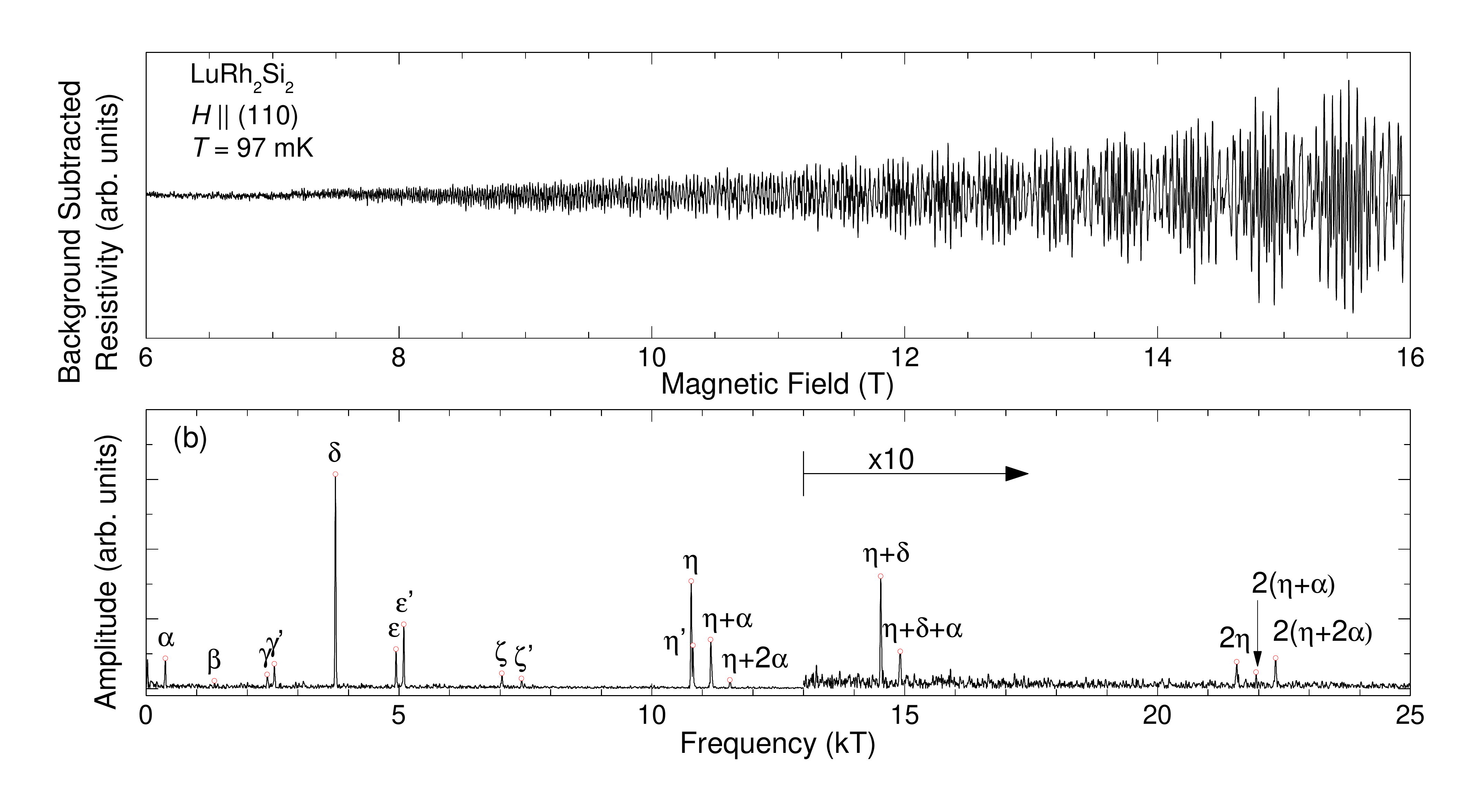}%
\caption{\SdH\ oscillations. (a) Oscillatory part of the resistivity between \unit{6}{\tesla} and \unit{16}{\tesla} for field along the (110) direction measured at \unit{100}{\milli\kelvin}. (b) Fourier transform indicating oscillation frequencies. Data above \unit{13}{\kT} are 10 times magnified for better view of the high frequency peaks.}
\label{fig:SdH}%
\end{figure}

Figure \ref{fig:SdH} shows a representative trace of the oscillatory part of the resistivity taken at \unit{95}{\mK} for fields between \unit{6}{\tesla} and \unit{16}{\tesla} with the field applied along the (110) direction. The power spectrum exhibits numerous oscillation frequencies. some of the frequencies have a signal to noise ratio of more than 100. For better visibility of the high frequency peaks, data in the Fourier spectrum above \Freq{13} are multiplied by a factor of 10. 19 frequencies are detected for this field orientation of which some are identified as harmonics and some arise from mixing of fundamental frequencies. 

The dominant frequencies are $\delta$ (\unit{3.7}{\kT}), $\epsilon^{\prime}$ (\unit{5.1}{\kT}), and $\eta$ (\unit{10.78}{\kT}). The beating seen in panel (a) is associated with the two frequencies $\eta$ and $\eta^{\prime}$ (\unit{10.81}{\kT}). In fact, the group of four frequencies around \unit{11}{\kT} comprises  two fundamental frequencies very close to each other ($\eta$ and $\eta^{\prime}$) whereas the others arise from mixing with once and twice the low frequency $\alpha$ (\Freq{0.38}). In addition, all frequencies above \Freq{11} were identified to arise from mixing of $\eta$ with $\delta$ and $\alpha$ in various combinations.

\begin{figure}%
\centering
\includegraphics[width=.7\textwidth]{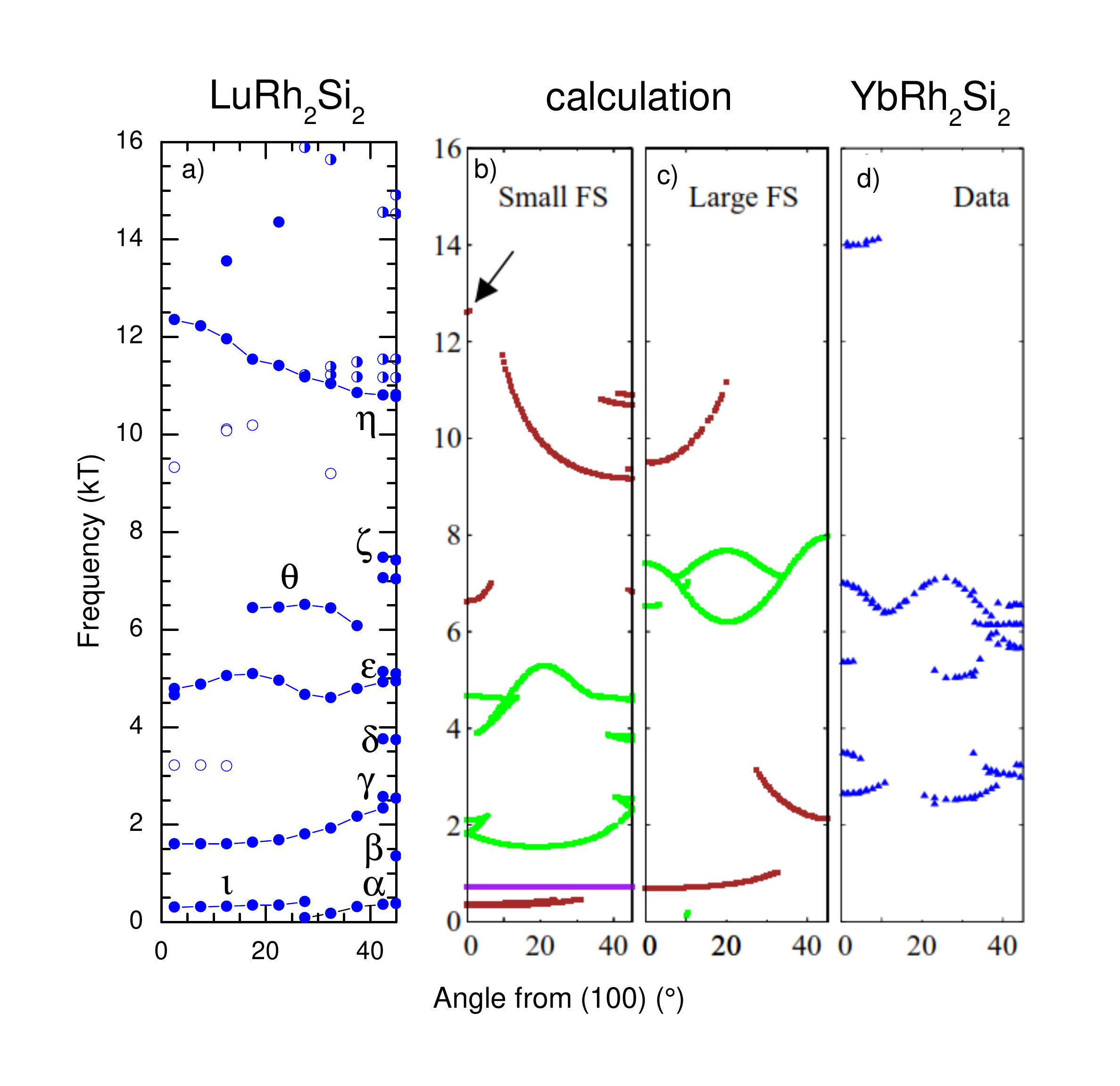}%
\caption{Angular dependence of  quantum oscillation frequencies in \LRS\ and \YRS. (a) Angular dependence of the SdH oscillation frequencies of our study on \LRS. Fundamental frequencies correspond to filled symbols whereas open symbols mark frequencies which have been identified as harmonics whereas half-filled symbols denote frequencies which were identified to arise from mixing of fundamental frequencies. Lines are guides to the eye, (b) and (c) reproduce the calculated angular dependences of quantum oscillation frequencies for \LRS\ and \YRS, respectively from Sutton et al. \cite{Sutton2010}. Green, red, and purple represent the ``doughnut'', ``jungle gym'', and ``pillbox'' Fermi surface sheet, respectively. (d) Shows the dHvA results on \YRS\ of Sutton et al. \cite{Sutton2010}. All panels have the same scales and show data for rotation of the magnetic field in the basal plane from (100) to (110) with the angle measured against (100).}
\label{fig:CompareYRS}%
\end{figure}

The angular dependence for field rotated in the basal plane is shown in Fig.~\ref{fig:CompareYRS} (a). Note that the trace depicted in Fig.~\ref{fig:SdH} corresponds to an angle of \unit{45}{\degree} in Fig.~\ref{fig:CompareYRS}(a). The main frequencies observed in Fig.~\ref{fig:SdH} obey distinct angular dependences. $\alpha$ is shifted to lower values and vanishes for angles of more than \unit{20}{\degree} away from (110). Here, a new frequency $\iota$ of similar size is seen all the way to field orientation parallel to (100) with small angular dependence. $\beta$ is not seen for field orientations away from the (110) direction. $\gamma$ is shifted to lower values and reaches \Freq{1.6} for field along (100). $\delta$ extends only a few degree with virtually no change of its value. $\epsilon$ obeys non-monotonic behaviour extending over the whole angular range. A new frequency $\theta$ emerges for intermediate angles, whereas $\zeta$ vanishes for more than \unit{5}{\degree} away from (110). $\eta$ shows a strong increase  from \Freq{10.78} to \Freq{12.36} as the field is rotated towards (100). The splitting into two frequencies, i.e., $\eta$ and $\eta^{\prime}$ was not observed away from (110). The mixing of frequencies discussed for the single trace in Fig.~\ref{fig:SdH} is confirmed in the rotation study. In summary, fundamental frequencies are observed in the range $\unit{0.08}{\kT} < F < \unit{12.5}{\kT}$.

These fundamental frequencies can be compared to the calculated  quantum oscillation frequencies of \YRS. Calculations were performed by P.M.C. Rourke et al. \cite{Rourke2008,Rourke2009} using the local density approximation including spin orbit interactions. This method is known to give good descriptions of the Fermi surface topology for non-correlated compounds (like \LRS) as well as for Ce-based heavy fermion systems. We note that for Yb-based heavy Fermion systems however, a renormalized band method might be necessary \cite{Zwicknagl2011}.  The comparison of calculated and observed quantum oscillation frequencies is done in Fig.~\ref{fig:CompareYRS} (b)-(d). 
First, Fig.~\ref{fig:CompareYRS} (b) depicts the predicted orbits for \LRS\ which was used to approximate the ``small'' Fermi surface configuration of \YRS. We note that these calculations employed the lattice parameters of \YRS\ ($a=4.010$ and $c = 9.841$) which differ only marginally from those of \LRS\ ($a = 4.007$ and $c = 9.839$). 
Due to these almost identical lattice parameters the comparison of our SdH measurements with the calculated orbits should hold to a quantitative level. However, there appears to be rather limited agreement between measured (Fig.~\ref{fig:CompareYRS} (a)) and calculated frequencies (Fig.~\ref{fig:CompareYRS} (b)). Best agreement is seen for the $\gamma$ orbit with both satisfactory agreement of the absolute value and the angular dependence to the predicted values of an orbit related to the so-called ``doughnut''  Fermi surface sheet. Rough agreement of $\epsilon$ with a second orbit of the ``doughnut'' sheet is seen although with significantly different angular dependence. For $\eta$ the agreement becomes unsatisfactory. It may be related to an orbit of the ``jungle gym'' although this should have a stronger angular dependence and is not predicted to extend to (100). $\iota$ might be mapped to another orbit of the ``jungle gym'' sheet whereas no agreement can be found for the $\alpha$ orbit with its strong angular dependence. The orbit predicted for the ``pillbox'' sheet expected at \Freq{1.4} is not observed. 

An agreement with the predictions for the ``large'' Fermi surface calculation (Fig.~\ref{fig:CompareYRS} (a)) is not expected as these were done for 
electronic configuration of \YRS. The comparison with the dHvA results on \YRS\ shows only remote similarities. The $\epsilon$ and $\theta$ orbits lie in a range where dHvA oscillations were observed but their angular dependence seems to be very different. No SdH oscillations were seen resembling the \Freq{14} orbit detected in \YRS, which was assigned to the ``jungle gym'' sheet of the ``small'' Fermi surface configuration. Although it is difficult to draw conclusions regarding \YRS\ from the absence of this orbit in the SdH measurements on \LRS, this seems to be significant. With the $\eta$, $\iota$, and probably the $\zeta$ orbit we detected key orbits assigned to the ``jungle gym'' surface. Consequently, the absence of an orbit similar to the \Freq{14} orbit of \YRS\ indicates that the assignment of this orbit to the ``small'' Fermi surface configuration is  questionable. In addition, the band structure calculations predict the \Freq{14} orbit to be limited to field orientations virtually parallel to the (100) direction, only. 
More detailed band structure calculations such as a renormalized treatment may be required for quantitative comparisons to the measured Fermi surface orbits of \YRS\ \cite{Sutton2010}.


%
%
\section{Conclusion}
In summary, a large set of SdH orbits has been observed for \LRS. The limited agreement with band structure calculations calls for further experimental and theoretical studies. We suggest to improve the band structure calculations by refining the underlying crystallographic parameters. Although the lattice parameters are known to high accuracy, the $z$-parameter specifying the position of the Si atoms has not yet been measured. In fact, preliminary results indicate a significant dependence of the Fermi surface structure on the $z$-parameter. 
Better understanding of the electronic structure of \LRS\ may help to improve the band structure calculations for \YRS\ which, in the case of the renormalized treatment, are based on the non-$f$ band structure. In addition, these calculations may provide a better understanding of the temperature and field dependence of the Hall effect in \LRS.

\ack
We acknowledge fruitful discussions with P.M.C. Rourke and G. Zwicknagl. This work was partially supported by the Alexander von Humboldt foundation, the Royal Society, and the EPSRC.

\section*{References}
\providecommand{\newblock}{}


\end{document}